\documentclass[aps,pra,twocolumn,superscriptaddress,showpacs,floatfix]{revtex4-1}
\usepackage{dcolumn,bm,textcomp}
\usepackage{mathtools}
\usepackage{amsfonts}
\usepackage{textcomp}
\usepackage{graphicx}
\usepackage{mathptmx}
\usepackage{xspace}
\usepackage{type1cm}
\usepackage{multirow}
\usepackage{textgreek}
\usepackage{natbib}
\usepackage{url}

\begin{document}
\title{New two-dimensional phase of tin chalcogenides: candidates for
high-performance thermoelectric materials}

\author{Baojuan Dong}
\affiliation{Shenyang National Laboratory for Materials Science,
Institute of Metal Research, Chinese Academy of Sciences, Shenyang 110016, China}
\affiliation{University of Chinese Academy of Sciences, Beijing 100049, China}
\affiliation{Skolkovo Institute of Science and Technology,
             Skolkovo Innovation Center,
             3 Nobel St., Moscow 143026, Russia}

\author{Zhenhai Wang}\thanks{physicswzh@gmail.com}
\affiliation{Skolkovo Institute of Science and Technology,
             Skolkovo Innovation Center,
             3 Nobel St., Moscow 143026, Russia}
\affiliation{School of Telecommunication and Information Engineering,
               Nanjing University of Posts and Telecommunications,
               Nanjing, Jiangsu 210003, China}
\affiliation{Emanuel Institute of Biochemical Physics RAS, 119334,
             4 Kosigin St, Moscow, Russia}

\author{Nguyen T. Hung}
\affiliation{Department of Physics, Tohoku University, Sendai 980-8578, Japan}

\author{Artem R. Oganov}
\affiliation{Skolkovo Institute of Science and Technology,
             Skolkovo Innovation Center,
             3 Nobel St., Moscow 143026, Russia}
\affiliation{Moscow Institute of Physics and Technology,
             9 Institutskiy Lane,Dolgoprudny City,
             Moscow Region 141700, Russia Federation}
\affiliation{International Center for Materials Discovery,
               Northwestern Polytechnical University,
               Xi'an, 710072, PR China}

\author{Teng Yang}\thanks{yangteng@imr.ac.cn}
\affiliation{Shenyang National Laboratory for Materials Science, Institute
of Metal Research, Chinese Academy of Sciences, Shenyang 110016, China}
\affiliation{University of Chinese Academy of Sciences, Beijing 100049, China}

\author{Riichiro Saito}
\affiliation{Department of Physics, Tohoku University, Sendai 980-8578, Japan}

\author{Zhidong Zhang}
\affiliation{Shenyang National Laboratory for Materials Science, Institute of
Metal Research, Chinese Academy of Sciences, Shenyang 110016, China}
\affiliation{University of Chinese Academy of Sciences, Beijing 100049, China}

\begin{abstract}
Tin-chalcogenides SnX (X = Te, Se and S) have been arousing research interest
due to their thermoelectric physical properties. The two-dimensional (2D)
counterparts, which are expected to enhance the property, nevertheless, have
not been fully explored because of many possible structures. Generating variable
composition of 2D Sn$_{1-x}$X$_{x}$ systems (X = Te, Se and S) has been performed
using global searching method based on evolutionary algorithm combining
with density functional calculations. A new hexagonal phase named by $\beta'$-SnX
is found by Universal Structure Predictor Evolutionary Xtallography (USPEX), and
the structural stability has been further checked by phonon
dispersion calculation and the elasticity criteria. The $\beta'$-SnTe is the
most stable among all possible 2D phases of SnTe including those experimentally
available phases. Further, $\beta'$ phases of SnSe and SnS are also found
energetically close to the most stable phases. High thermoelectronic (TE) performance
has been achieved in the $\beta'$-SnX phases, which have dimensionless figure of
merit (ZT) as high as $\sim$0.96 to 3.81 for SnTe, $\sim$0.93 to 2.51 for SnSe and
$\sim$1.19 to 3.18 for SnS at temperature ranging from 300 K to 900 K with practically
attainable carrier concentration of 5$\times$10$^{12}$ cm$^{-2}$. The high TE
performance is resulted from a high power factor
which is attributed to the quantum confinement of 2D materials and the band convergence
near Fermi level, as well as low thermal conductivity mainly from both low elastic
constants due to weak inter-Sn bonding strength and strong lattice anharmonicity.
\end{abstract}
\date{\today}

\maketitle

\section{Introduction}
Group IV-VI alloys have been intensively studied with its many physical properties
including ferroelectricity~\cite{Chang16}, topological insulator~\cite{Tanaka12}
and, in particular, thermoelectricity~\cite{Zhao14,Heremans08,Li15}. Thermoelectric
(TE) materials, which directly convert waste heat into electricity, have drawn an
attention in the last few decades. The conversion efficiency of TE materials can be
evaluated by the dimensionless figure of merit $ZT$ (= $\frac{\sigma S^2}{\kappa}T$,
in which $\sigma$, $S$, $\kappa$ and $T$ represent the electrical conductivity,
Seebeck coefficient, thermal conductivity and temperature, respectively). Among the
group IV-VI alloys, tin and lead chalcogenides~\cite{Zhao14,Heremans08,Lee14,Li15}
have been attracting increasing interest in thermoelectric community with the structural
and electronic structural anisotropy and intrinsic lattice anharmonicity~\cite{Lee14,Li15}.
Lattice anharmonicity helps to suppress thermal conductivity, while anisotropy is
related to the confinement effect which has proved to be efficient in improving
thermoelectric performance according to Hicks-Dresselhaus theory~\cite{Dresselhaus1993,Dresselhaus1993b}
if the confinement length is smaller than thermal de Broglie length~\cite{Nguyen16}.

\begin{figure*}[t]
\includegraphics[width=0.8\linewidth]{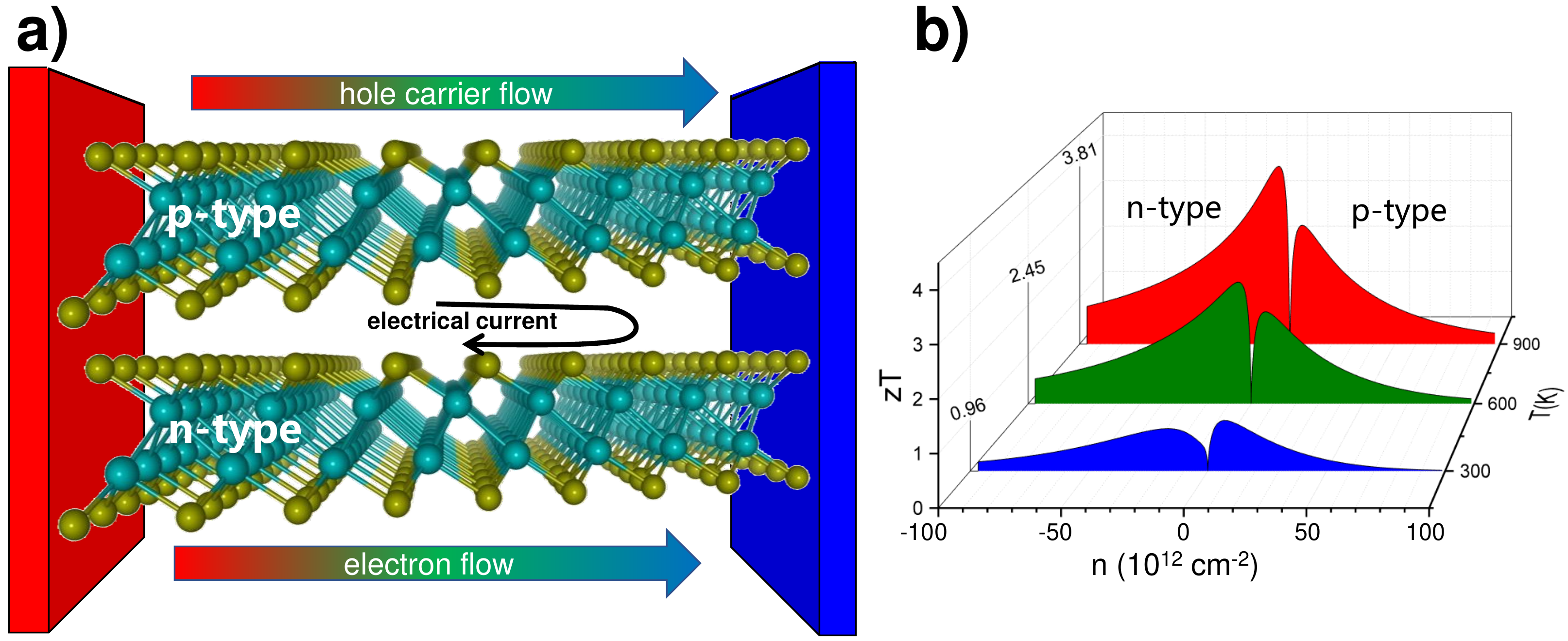}
\caption{\textbf{Promising thermoelectric properties of new structural phase of SnTe}.
(a) Schematics of $\beta'$-SnTe for thermoelectrics, (b) temperature and carrier
concentration dependent dimensionless figure of merit $ZT$. The values (0.96, 2.45, 3.81)
on the left panel are the peak $ZT$ values of $\beta'$-SnTe at 300K, 600K, 900K,
respectively.}
\label{fig:fig1}
\end{figure*}

With the development of exfoliation and synthesis method, many two-dimensional (2D)
van der Waals materials including graphene, black phosphorene (BP), transition metal
dichalcogenides (TMDs) and tin chalcogenide (SnX) has been
synthesized~\cite{Novoselov04, Tony2010, Liu14, Wang2013}. The exfoliated semiconducting
monolayer BP and SnX have shown much improved thermoelectric performance ($ZT$ $\sim$
2.5~\cite{Fei14} and 2.63~\cite{Han2017} at 500K and 700K, respectively) with respect
to the bulk counterparts. Therefore, it is reasonable to focus our attention on 2D
counterpart of tin chalcogenides, which is promising candidates for high-performance
TE materials. So far, there has been few studies on the TE properties of the 2D tin
chalcogenides, except for some limited theoretical calculations~\cite{Li2017,Han2017}.
Moreover, the 2D forms may exhibit many different structures from the bulk counterpart,
especially for the non-layered bulks~\cite{Robinson2016, Guy2010}. Thus, it is naturally
curious to investigate theoretically the most stable structure among all the possible
forms and explore the potential TE performance.

In this work, using ''Universal Structure Predictor Evolutionary Xtallography'' USPEX
method~\cite{USPEX1,USPEX2,Oganov10,Lyakhov10}, we survey all possible Tin-chalcogenide
2D phases. A new hexagonal SnX (X= Te, Se, S) phase, which is named $\beta'$ phase and
shown in Fig.~\ref{fig:fig1}(a), has been found by using USPEX. The $\beta'$-SnX have
been checked to be thermodynamically stable. Owing to the low lattice thermal conductivity
$\kappa$$_{l}$ and high $\sigma$, as explained in the following section, high thermoelectric
performance is achieved in the $\beta'$-SnX phases. For example, as seen from Fig.~\ref{fig:fig1}(b),
$ZT$ of $\beta'$-SnTe at a carrier concentration around a few 10$^{12}$ cm$^{-2}$ can
be obtained up to 2.45 and 3.81 at 600 and 900 K, respectively.

The paper is organized as follows. We briefly introduce the computational methods in II.
In III we shows the main results of the $\beta'$-SnX, including the structural stability
in III-A, the thermal transport properties in III-B, and thermoelectric properties in III-C.
Finally we draw a conclusion in IV.

\section{Method}
The structure search of 2D tin chalcogenides is performed by USPEX~\cite{USPEX1,USPEX2,Oganov10,Lyakhov10}
combined with Vienna $\emph{ab initio}$ simulation package (VASP)~\cite{VASP}. In our
variable-composition USPEX calculations, the thickness of 2D crystals is restricted in range
of 0-6 \AA, the total number of atoms is set to be 2-12, while 80 layer groups are chosen for
the symmetry in generation of initial 2D structures. Total energy is calculated
within the framework of Projector Augmented Wave (PAW) method~\cite{PAWPseudo}. Generalized
gradient approximation (GGA)~\cite{PBE} is used to treat the electronic exchange correlation
interaction. More details on the parameters can be referred to the supplementary information~\cite{supplementary}.

Electronic transport properties are calculated by solving the semi-classical Boltzmann transport
equation within the constant relaxation time approximation as implemented in the BoltZTraP
package~\cite{BoltZTrap}. Since there is no experimental data of electrical conductivity available
for the new $\beta$'-phases to evaluate the relaxation time $\tau$, $\tau$ is estimated based on
carrier mobility $\mu$, for example, $\tau = \frac{m^{*}\mu}{e}$, $m^{*}$ is the effective mass
of carrier, carrier mobility $\mu$ is calculated based on the deformation potential
theory~\cite{Shockley1950,Ji2014,Zhang2014,Nuo2015,Su2016,Sun2016}. The calculated $\tau$ at room
temperature varies from a few tens to a few hundred femtoseconds (10$^{-15}$ s), which has the same
order of magnitude as the calculated values in other 2D materials~\cite{Shafique17, Jiang17}. More 
details on how to get the carrier mobility $\mu$ and relaxation time $\tau$ can be found in the supplementary 
information~\cite{supplementary}. 
Although the relaxation time of electron depends on the Fermi energy, we adopted the constant
relaxation approximation for simplicity. In order to calculate the relaxation time by first principles
calculations, we should consider electron-phonon interaction and phonon-phonon interaction for
estimating the relaxation time of an electron and a phonon in the same level of approximation,
which should be a future problem.
Phonon dispersion relation is calculated by Phonopy package~\cite{phonopy}. The $\kappa$$_{l}$ is
evaluated by phonon lifetime, which is self-consistently calculated in the ShengBTE
package~\cite{ShengBTE}. The second-order harmonic interatomic force constants (IFCs)
are calculated within the Phonopy package, and the third-order anharmonic IFCs
are evaluated by using 3$\times$2$\times$1 supercell and up to the fifth-nearest neighbors
considered by ShengBTE.

\begin{figure*}[t]
\includegraphics[width=0.70\linewidth]{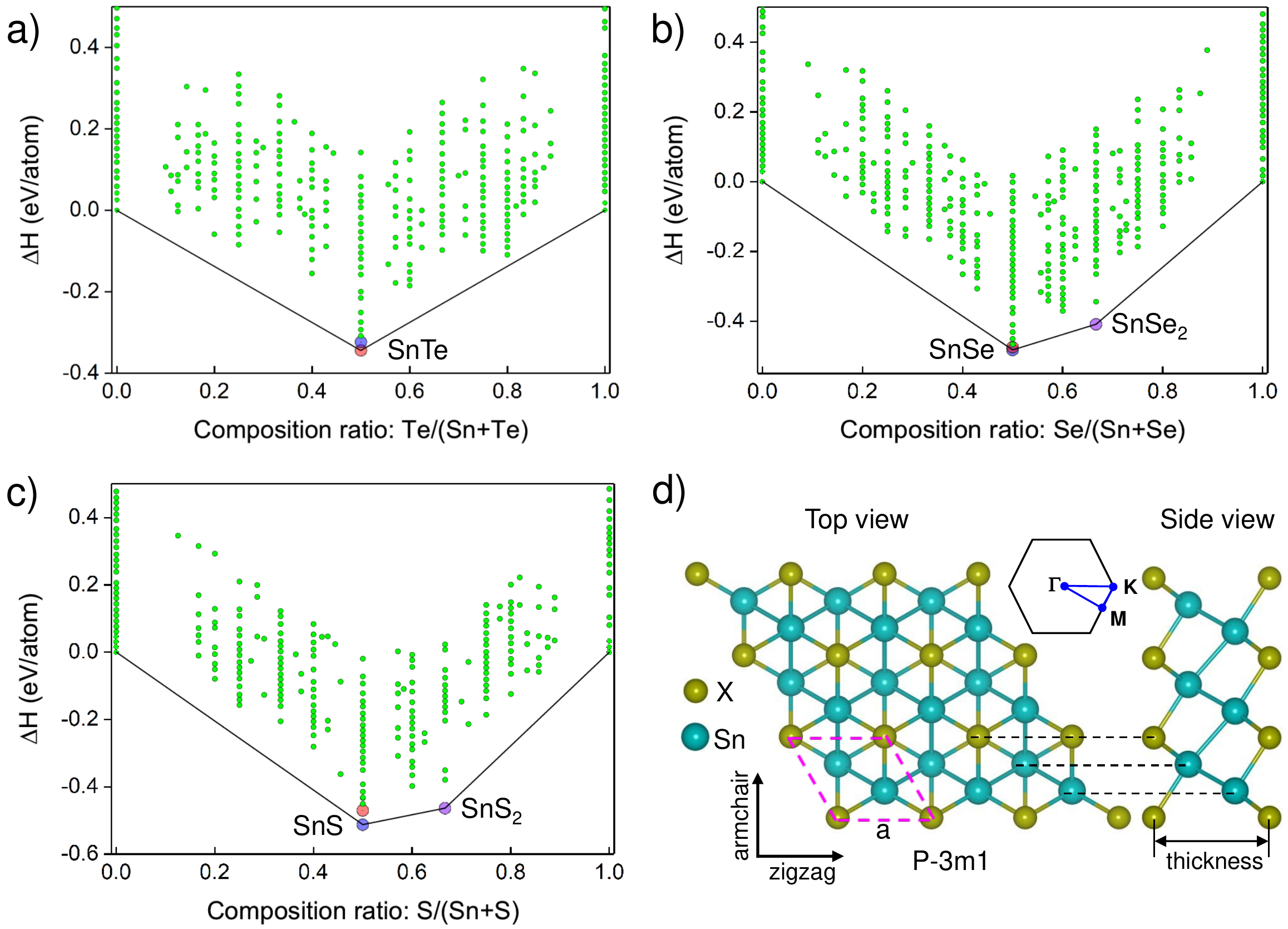}
\caption{\textbf{Convex hull of Sn-X (X = Se, S and Te) materials searched by USPEX and atomic
structure of $\beta'$-SnX}. USPEX-predicted formation enthalpy $\Delta$H of 2D bi-element
structures with different stoichiometries between (a) Sn and Se, (b) Sn and S, and (c) Sn and
Te. The blue and purple dots represent the stable structural phases experimentally observed,
and the red dots represent the new $\beta'$-phase of SnX. (d) Both top and side views of atomic
structure for $\beta'$-SnX, where light-blue and dark-yellow represent Sn and X (X = Te, Se, S),
respectively. The unit cell is marked by the pink dash lines and the first Brillouin Zone is
shown. Armchair, zigzag and thickness directions are indicated by arrows.}
\label{fig:fig2}
\end{figure*}

\begin{table*}[t]
\centering
\caption{Structural and mechanical parameters for the $\beta'$-SnX. Here a is the lattice
constant, $b_{Sn-Sn}$ ($b_{Sn-X}$) is the bond length for Sn and Sn(X),  and thickness for
2D SnX is the vertical distance between the two outermost X atoms in the unit of angstrom,
which are shown in Fig. 1. In-plane Young and shear module in the unit of $N m^{-1}$ are listed. }
\label{table1}
\begin{tabular}{ p{1.5cm}<{\centering}  p{1.5cm}<{\centering}  p{1.5cm}<{\centering}
p{1.5cm}<{\centering} p{2cm}<{\centering} p{2.5cm}<{\centering} p{2.5cm}<{\centering} p{2cm}<{\centering}}
\hline
& a  &  $b_{Sn-Sn}$   &  $b_{Sn-X}$  & thickness & Young's modulus & shear modulus & Poisson ratio  \\
&  (\AA)  &  (\AA)  &   (\AA) & (\AA) &  ($N m^{-1}$) & ($N m^{-1}$) &  \\ \hline
SnTe &  4.34 & 3.36 &  2.97  & 5.39 & 47.15 & 18.42 & 0.28\\
SnSe &  4.09 & 3.37 &  2.76  & 5.26 & 45.77 & 17.17 & 0.33\\
SnS  &  3.95 & 3.38 &  2.64  & 5.12 & 45.32 & 16.53 & 0.37\\ \hline
\end{tabular}
\end{table*}

\begin{figure*}[t]
\includegraphics[width=0.8\linewidth]{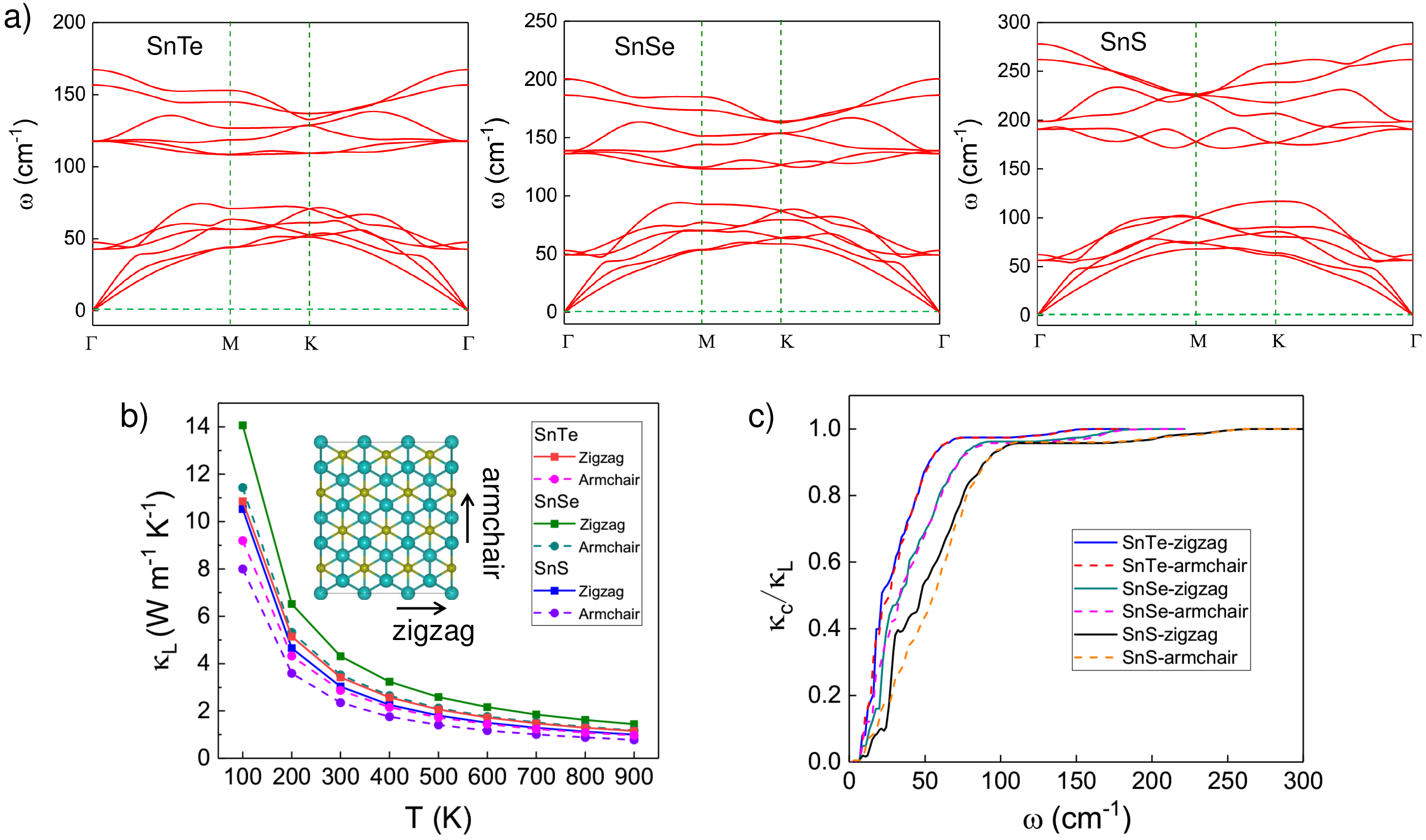}
\caption{\textbf{Lattice thermal properties of $\beta'$-SnX}. (a) Phonon dispersion relation,
(b) lattice thermal conductivity $\kappa_L$, and (c) normalized cumulative thermal conductivity
$\kappa_c$/$\kappa_L$ as a function of phonon frequency.}
\label{fig:fig3}
\end{figure*}

\section{Results and Discussions}

To understand such a high thermoelectric performance in the $\beta'$-SnX, we investigate the
structural, thermal and electronic transport properties of the $\beta'$-SnX systems as follows.

\subsection{Structure and stability of $\beta'$-SnX}
First, we show the results of global search on 2D structures of tin chalcogenides Sn$_{1-x}$X$_{x}$
(X = Te, Se, S). In Fig.~\ref{fig:fig2}, we show the formation enthalpy $\Delta$H (defined in Eq.(1) in
supplementary) of tin chalcogenide 2D systems as a function of chalcogenide composition in the
variable-composition convex hulls as predicted by USPEX. In the convex hull, the zero line connects
two points at $\Delta$H = 0 for the most stable 2D elementary structures of Sn and chalcogenide as
predicted by USPEX. 2D Sn$_{1-x}$X$_{x}$ is more stable than the reactant materials only when $\Delta$H
is below the zero line. Out of more than 2600 structures generated, we show only those with $\Delta$H
lower than 0.5 eV/atom. The two most stable structures of Sn-Te as highlighted in blue and red dots
in the convex hull in Fig.~\ref{fig:fig2}(a) are the commonly observed puckered orthorhombic SnTe
phase (Fig.1S(a) in the supplementary)~\cite{Chang16} and the new $\beta'$ SnTe phase, respectively.
$\Delta$H of the $\beta'$ SnTe is lower by 19 meV/atom than the puckered orthorhombic SnTe phase.
The $\beta'$ SnTe has actually been proposed to be a stable semiconductor by Sa~\cite{Sa16} and
Zhang et al.~\cite{Chen17}. Here we substantiate structural stability of the $\beta'$ phase in a
convex hull with all possible stoichiometries considered.

The $\beta'$ phase of both Sn-Se and Sn-S has also been obtained close to the convex hulls, as shown
in red dots in Fig.~\ref{fig:fig2}(b,c). However, the $\beta'$ phase is less stable than the puckered
orthorhombic phase~\cite{Li2013}, with $\Delta$H slightly higher by 8 meV/atom for SnSe and 42 meV/atom
for SnS. Additionally, octahedral 1T phases of both SnSe$_2$ and SnS$_2$, the former of which has been
synthesized by experiment~\cite{zhou2015}, is found to be stable in the convex hull, as shown in purple
dots in Fig.~\ref{fig:fig2}(b,c). In the current paper, we will focus only on the $\beta'$ phases.

All the $\beta'$ phases have $P\bar{3}m1$ symmetry (space group $\sharp$164), with the optimized
atomic structure and lattice parameters of $\beta'$-SnX (X = Te, Se, S) shown in Fig.~\ref{fig:fig2}(d)
and Table~\ref{table1}, respectively. From Fig.~\ref{fig:fig2}(d), the $\beta'$ structure can be viewed
as a buckled hexagonal lattice of Sn with two X atoms (one up and one down) at the center of hexagon.
Or as shown in Fig.1S(b) in supplementary, it can be considered as two stacked $\beta$-SnTe monolayers,
one of which takes a series of symmetry operations (inversion + glide) to get the second layer, which
makes the $\beta'$ phase distinct from and more stable than AB-stacked $\beta$-bilayer in which a translation
symmetry exists between two $\beta$ monolayers. The relative stability of the $\beta'$-SnTe phase over
the AB-stacked $\beta$-bilayer is analyzed in more details in supplementary~\cite{supplementary}.

To check the stability of the $\beta'$-SnTe, we calculated the phonon dispersion relation of the $\beta'$
phase of SnX, as shown in Fig.~\ref{fig:fig3}(a). No imaginary phonon frequencies are found near the
$\Gamma$ point, showing that the $\beta'$ phases are dynamically stable. And the stability is also
checked by the elastic parameters in Table~\ref{table1} from the standard criteria of elastic
stability~\cite{Mouhat2014,Born40,Zhou08}. In fact, the necessary condition for stable 2D materials is
that all elastic constant C$_{ij}$ should be positive~\cite{Wang1999}, which is satisfied in our
$\beta'$-SnX materials (more details in supplementary~\cite{supplementary}).

\begin{figure*}[t]
\includegraphics[width=0.85\linewidth]{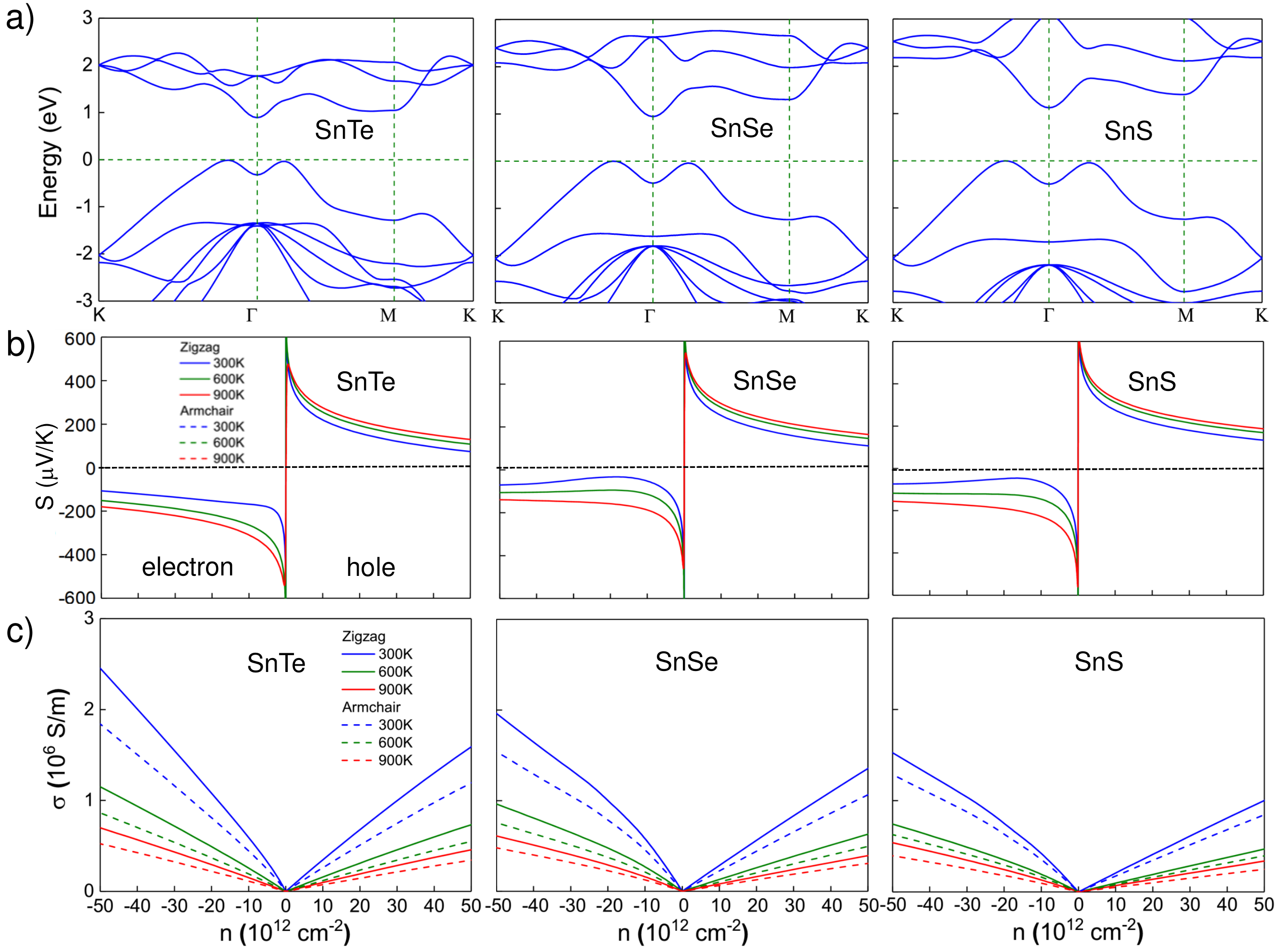}
\caption{\textbf{Electrical transport properties of $\beta'$-SnX}. (a) Electronic band structure. (b)
Seebeck coefficients S, and (c) electrical conductivity $\sigma$ of SnX for zigzag and armchair directions,
as a function of carrier density for T = 300, 600 and 900 K.}
\label{fig:fig4}
\end{figure*}

\begin{table*}[t]
\centering
\caption{Carrier mobility at 300K and effective mass for SnX. The effective mass is in units of
electron mass m$_{0}$ (9.11$\times$10$^{-32}$ $kg$). The method in the supplementary gives more details
on calculating carrier mobility $\mu$.~\cite{supplementary}}
\label{table2}
\begin{tabular}{ p{3cm}<{\centering} p{1.5cm}<{\centering} p{1.5cm}<{\centering}  p{1.5cm}<{\centering}
p{1.5cm}<{\centering} p{1.5cm}<{\centering} p{1.5cm}<{\centering} p{1.5cm}<{\centering}}
\hline
& & \multicolumn{2}{c}{SnTe}  &  \multicolumn{2}{c}{SnSe}  & \multicolumn{2}{c}{SnS}   \\
& &  hole &  electron &  hole  &  electron &  hole & electron \\ \hline
carrier mobility $\mu$ & zigzag    & 1364  & 1112  & 1275  &  853  & 1220  & 764 \\
($cm^{2} V^{-1} s^{-1}$)&armchair &  576  &  834  & 579   &  694  &  468  & 660 \\
effective mass $m^*$ & zigzag      & 0.213 & 0.144 & 0.228 & 0.169 & 0.363 & 0.218  \\
($m_{0}$)            & armchair    & 0.227 & 0.144 & 0.228 & 0.163 & 0.363 & 0.212 \\ \hline
\end{tabular}
\end{table*}

\subsection{Thermal transport properties of $\beta'$-SnX}
Thermoelectric properties consist of thermal and electronic transport properties. We first study the
thermal transport properties of $\beta'$-SnX. Since the thermal transport of lattice is related to
the mechanical properties, let us discuss the mechanical properties firstly. The Young's modulus of
the $\beta'$-SnX (less than 50 N/m) as shown in Table~\ref{table1} are much smaller than other 2D
materials like graphene ($\sim$345 N/m) and phosphorene ($\sim$23-92 N/m)~\cite{Kou15}. Shear modulus
of the $\beta'$-SnX are found less than 20 N/m. From the phonon dispersion in Fig.~\ref{fig:fig3}(a),
we can see anti-crossing of the phonon dispersion between low-frequency optical vibration modes with
acoustic phonon modes for the $\beta'$-SnX. These optical phonon modes correspond to two-fold degenerate
in-plane shearing modes and out-of-plane breathing mode. Interestingly, their vibrational frequency
are almost independent of materials at around 50 cm$^{-1}$ at $\Gamma$ point, which is closely related
to their similar and low values of the shear modulus as given in Table~\ref{table1}. In Fig.~\ref{fig:fig3}(b),
the calculated $\kappa_L$ is plotted as a function of $T$ for armchair and zigzag directions. For
example, $\kappa_L$ of $\beta'$-SnTe at 300 $K$ along armchair direction is as low as 2.87 $Wm^{-1}K^{-1}$.
A typical T dependence of $\kappa_L$ ($\kappa_L \sim 1/T$)
reveals that the Umklapp process in the phonon scattering is essential for the temperature range that
we studied. In Fig.~\ref{fig:fig3}(c), we show the normalized $\kappa_L$ by cumulative thermal conductivity
$\kappa_{c}$ as a function of frequency $\omega$ at room temperature. $\kappa_{c}$ is the value
of $\kappa_{L}$ when only phonons with mean free paths below a threshold are considered~\cite{ShengBTE}.
Over 90$\%$ of the $\kappa_L$ is contributed by phonon modes with frequency below 80 cm$^{-1}$ for SnTe,
in which the three acoustic modes and the three low-frequency optical modes contribute to $\kappa_L$.

The low lattice thermal conductivity of the $\beta'$-SnX arises not only from low elastic constants due
to weak Sn-Sn bonding strength, but also from strong lattice anharmonicity. In the low-frequency region,
Fig.S3(a) in supplementary shows that the anharmonic scattering dominates the phonon-phonon interactions
(PPIs) by comparing the anharmonic three-phonon scattering rates (ASRs) and isotropic scattering rates
(ISRs). These ASRs are mainly contributed by phonon absorption process (ASRs+). Among the three $\beta'$-SnXs,
the strongest ASRs are found in SnS, which corresponds to the lowest $\kappa$$_{L}$. It is worth noting
that the highest value of ASRs are located between around 50 and 100 cm$^{-1}$, where acoustic and three
low-frequency optical modes are mixed to one another. Thus we expect that the inter-band scattering between
the acoustic and optical modes are associated with the large ASRs.

\begin{figure*}[t]
\includegraphics[width=0.75\linewidth]{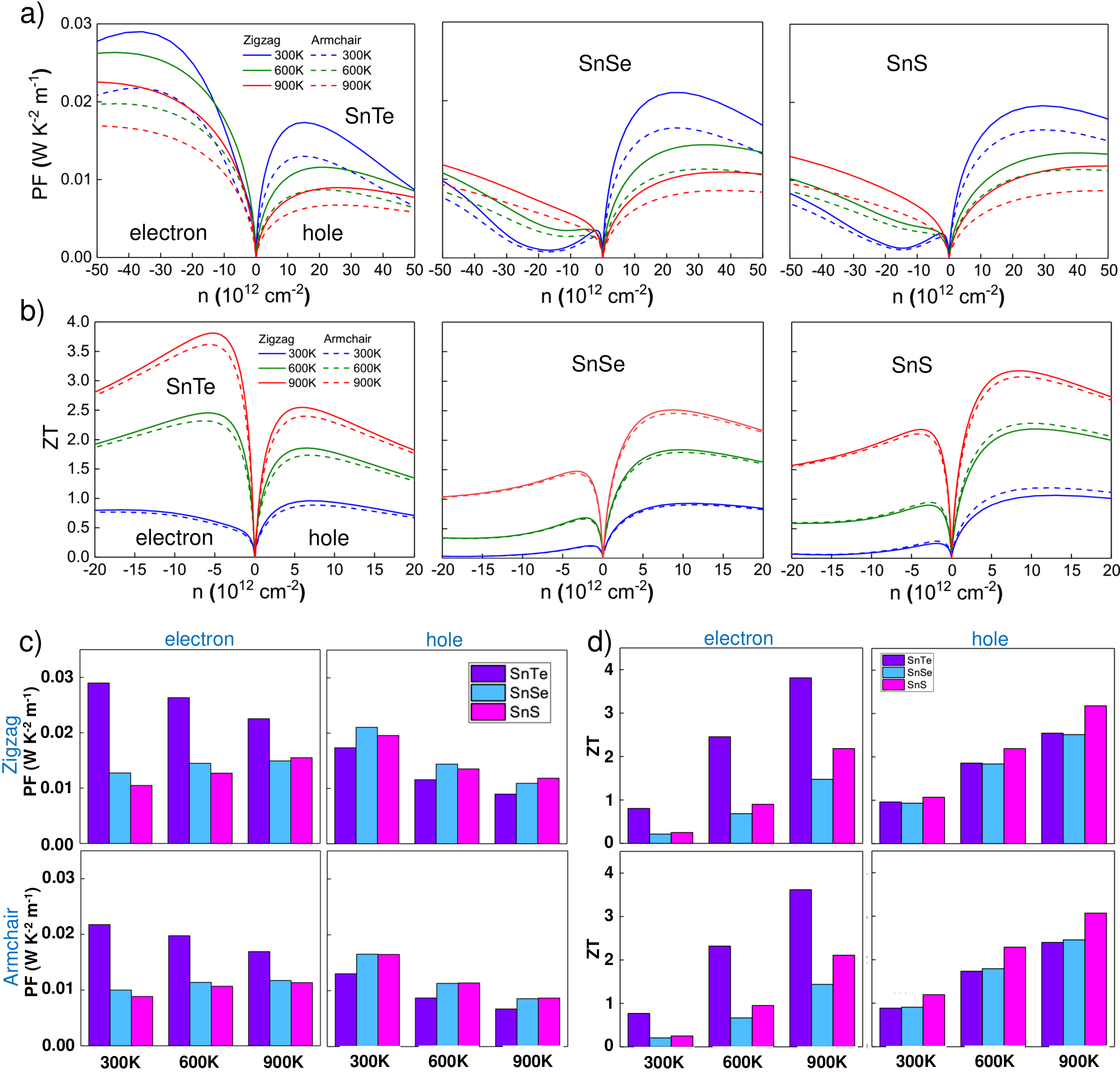}
\caption{\textbf{Thermoelectric performance for $\beta'$-SnX}. (a,c) Power factor (PF)
and (b,d) Figure of merit $ZT$ of SnX as a function of doping level $n$ at different
temperature. Here $n$ is the electron (negative) or hole (positive) doping per unit
surface area for 2D SnX. Blue, green and red color represents 300K, 600K and 900K,
and solid and dash line represents zigzag and armchair direction which is shown in
Fig.\ref{fig:fig2}. The maximal PF and $ZT$ at optimal doping level as a function of
temperature and crystal direction are shown in (c) and (d).}
\label{fig:fig5}
\end{figure*}

\subsection{Thermoelectric properties of $\beta'$-SnX}

Based on the calculated results, we will discuss the thermoelectric properties of
$\beta'$-SnX.

\subsubsection{Seebeck coefficient}

The calculated structural stability and low thermal conductivity suggest that
$\beta'$-SnX can be considered suitable for thermoelectric applications.
To unveil its potential for energy conversion between heat and electricity,
we look into the relevant electronic band structure and electrical properties,
both of which reinforce its capacity for such an application. In Fig.~\ref{fig:fig4}(a),
we show electronic band structures of $\beta'$-SnXs. Indirect band gaps of $\beta'$-SnXs
exist near the zone center. The value of energy gaps are around 1.0 eV, which
are independent of chalcogenide atoms. It is noted that the conduction bands
have been upshifted to fit the band gap obtained by hybrid functional
calculations~\cite{HSE06}, which usually give a more reliable band gap size.
According to E$_g$ $\sim$ 10 K$_\textrm{B}$T$_\textrm{opt}$ rule~\cite{Mahan}, the optimal working
temperature for thermoelectric applications of such materials should be around 1000 K.

Thermoelectric properties are closely related to electronic band structure. For all
$\beta'$-SnXs, we found the following features in the electronic bands: (1) Band dispersions
of valence band maximum (VBM) and conduction band minimum (CBM) along both $\Gamma$K
and $\Gamma$M directions are quite similar, which corresponds to a similar effective
mass along both the zigzag and armchair directions as given in Table~\ref{table2}.
According to Cutler et al.~\cite{Cutler64} and Snyder et al.~\cite{Jeffrey18}, for
a parabolic band within the energy-independent scattering approximation, the Seebeck
coefficient takes the form of S = $\frac{8 \pi^2 k^2_B m^* T}{3 e h^2} (\frac{\pi}{3n})^{2/3}$,
where m$^*$ is the effective mass of the carrier and $n$ is the carrier concentration.
From this formula, similar to the effective mass m$^*$, one expects no directional
dependence of the Seebeck coefficient, as shown in Fig.~\ref{fig:fig4}(b), in which
dashed lines are completely overlapped by solid lines. (2) A usual quadratic
dispersion relation appears for the
carriers at the CBM, while a quartic band dispersion (E$_k$ $\sim$ k$^4$) is found at
the VBM, which usually brings about flat bands near Fermi level. Thus constant
electronic density of states (DOS) appears near CBM, while a van-Hove DOS singularity
divergence appears near VBM~\cite{Seixas16}, as shown in Fig.4S in supplementary.
Such a difference of DOS between VBM and CBM explains why the effective mass of holes is
larger than that of electrons, as is listed in Table~\ref{table2}.

In Fig.~\ref{fig:fig4}(b) and (c), we show Seebeck coefficient and electrical conductivity.
For $\beta'$-SnSe and $\beta'$-SnS, the Seebeck coefficient of hole carriers is larger than
that of electron; while the electrical conductivity $\sigma$ of hole is smaller than that
of electron, which is expected for a parabolic band within energy-independent scattering
approximation~\cite{Jeffrey18}. However, an opposite trend is found in $\beta'$-SnTe that
Seebeck coefficient of electron is higher than that of hole, which is due to the convergence
of conduction band minimum at $\Gamma$ point with flat band edge at $M$ point. Such type
of band convergence is much advantageous for an enhancement of Seebeck coefficient~\cite{Nguyen18}.

\subsubsection{Electrical conductivity}
In Fig.~\ref{fig:fig4}(c), we plot the calculated electrical conductivity $\sigma$ as a
function of carrier concentration $n$ for zigzag and armchair directions. Decent
electrical conductivity $\sigma$ as high as a few 10$^6$ S/m at room temperature is
obtained for the three $\beta'$-SnXs. Differences of $\sigma$ for different
materials along different directions can be understood
from the carrier mobility $\mu$ and the effective mass m$^*$ as is listed in Table~\ref{table2}.
From Table~\ref{table2}, we can point out that (1) the effective mass m$^*$ depends not
on crystal direction, but on carrier type, for example, m$^*_{h}$ $>$ m$^*_{e}$;
(2) carrier mobility $\mu$ along the zigzag direction is larger than that along
the armchair direction, due to a smaller deformation potential along the zigzag
direction than the armchair direction. All these features lead to a preference of zigzag
over armchair direction and electron over hole carrier for optimal electrical
conductivity $\sigma$, which is indicated by the comparison between solid (zigzag)
and dashed (armchair) lines as is shown in Fig.~\ref{fig:fig4}(c). Moreover, $\sigma$
decreases with increasing temperature, which is associated with the intrinsic phonon
scattering mechanism. The electrical thermal conductivity $\kappa_{e}$ is
also calculated based on the Boltzmann transport theory, as given in
Fig.5S, and fits the Wiedemann-Franz law in combination with $\sigma$.

\subsubsection{Power factor and figure of merit}

With Seebeck coefficient, electrical conductivity and thermal conductivity
available, we finally evaluate power factor (PF) and dimensionless figure
of merit (ZT). In Fig.~\ref{fig:fig5}(a) and (b), we plot the dependence
of carrier type, crystalline direction and temperature for PF and ZT.
The optimal PF and $ZT$ are shown in Fig.~\ref{fig:fig5}(c,d). As seen from
Fig.~\ref{fig:fig5}(a), PF remains as high as 0.01 WK$^{-2}$m$^{-1}$ or
above in a wide range of temperature at carrier concentration from 10$^{12}$
to 10$^{13}$ cm$^{-2}$. Because of the high PF and relatively low $\kappa$,
it is no surprise to observe quite promising value of $ZT$ in $\beta'$-SnX.
From Fig.~\ref{fig:fig5}(b), all $ZT$ of $\beta'$-SnX show above 1.0 at 900 K
in the interested doping range ($|n|$ $<$ 8 $\times$ 10$^{13}$ cm$^{-2}$).
$ZT$ of $\beta'$-SnTe can even go above 2.0 at 600K, which makes the material
very competitive against the present commercialized thermoelectric materials.
From Fig.~\ref{fig:fig5}(a, b), both PF and $ZT$ are larger for hole than for
electron in $\beta'$-SnS and $\beta'$-SnSe, mainly due to the smaller Seebeck
coefficient of electron than hole. As for $\beta'$-SnTe, we get a better
thermoelectric performance of electron than hole, which is due to a large
S and PF from the concept of `band convergence'~\cite{Nguyen18,Pei11} at CBM
concurrent with decent electrical conductivity from the smaller effective mass
of electron than that of hole. In Ref.~\cite{Nguyen18}, the dependence of optimal
PF$^{\textrm{opt}}$ on $\Delta$E for a generic systems with band convergence
is given quantitatively within two-band model, with $\Delta$E defined as valley
splitting energy. PF$^{\textrm{opt}}$ decreases exponentially with increasing
$\Delta$E within a few k$_B$T. In our case, $\Delta$E is the energy difference
of the CBMs between the K and M points in Fig.~\ref{fig:fig4}(a), $\Delta$E is
0.15, 0.36 and 0.28 eV for $\beta'$ SnTe, SnSe and SnS, respectively, which
explains why a much bigger PF$^{\textrm{opt}}$ of the n-type $\beta'$ SnTe is
obtained than that of the n-type $\beta'$ SnSe and SnS.

In Fig.~\ref{fig:fig5}(c,d), we show the optimal values of PF and $ZT$ for two
types of carriers along armchair and zigzag directions at T = 300, 600 and
900 K. It is more clear to see in Fig.~\ref{fig:fig5}(a,b) that n-type
$\beta'$-SnTe has a much better thermoelectric performance than $\beta'$-SnS
and $\beta'$-SnSe, while p-type $\beta'$-SnX shows very decent performance but
no obvious difference of PF and $ZT$ from SnS to SnTe.

It should be pointed out that we expect some discrepancy between theoretical
and experimental values. For our estimation, there are following reasons for
discrepancy: 1) the constant relaxation time approximation was used for
electronic transport properties, where the real relaxation time may vary
with the carrier concentration; 2) only isotopic and three-phonon scattering
was considered here for $\kappa_{L}$. The constant relaxation time approximation
may overestimate the $\sigma$. However, $\kappa_{L}$ may also be overestimated
without considering enough scattering rates coming from the impurity, defect,
grain boundary and dislocation and so on. Considering that the two parameters
$\kappa_{L}$ and $\sigma$ are both overestimated, the deviation of TE performance
may be alleviated in part by the two effects. Therefore, our estimated TE
performance may give a reasonable agreement with the experimental values.

Finally due to confinement effect for 2D system~\cite{Dresselhaus1993,Dresselhaus1993b},
it is important to evaluate the PF enhancement factor f$_E$~\cite{Nguyen16}, which is
defined as f$_E$ = $(\frac{L}{\Lambda})^{D-3}$, where L is the spatial confinement
length and $\Lambda$ (= $\sqrt{\frac{2\pi\hbar^2}{k_BTm^*}}$) is the so-called thermal
de Broglie wavelength and D (= 1 or 2) is the dimension. Here we consider the PF
enhancement from 3D to 2D. L is taken from the interlayer distance in 3D counterparts.
Values of $\Lambda$, L and F$_E$ for $\beta'$-SnX are given in Table III in supplementary.
Take n-type $\beta'$-SnTe as an example, we find that $\Lambda$ $\sim$ 11.35 nm, L
$\sim$ 0.82 nm, which makes f$_E$ $\sim$ 13.76, about one order of magnitude from 3D
to 2D, revealing that thermoelectric behavior of $\beta'$-SnX in 2D form is much enhanced
upon its 3D counterpart.

In summary, for the new $\beta'$ phase of SnX, the decent thermoelectric properties beyond
traditional thermoelectric materials occur because of the following reasons: (1) The low
dimensional structure with high elastic and dynamic stability, which shows substantial
enhancement of power factor upon the bulk phase due to the larger thermal de Broglie length
$\Lambda$. (2) The low shear modulus within the layer giving rise to an ultralow
frequency of the shearing mode which can couple very effectively with the acoustic phonon
mode to greatly suppress the lattice thermal conductivity. (3) The convergence of electronic
bands at the valence and conduction band edges. All the above factors appear concurrently
and coherently to lead to the good thermoelectric performance.

\bigskip

\section{conclusion}
In this paper, we found new-phase SnX ($\beta'$ phase) which is suitable for thermoelectric
application by combining \emph{ab initio} density functional theory with genetic algorithm
and semi-classical Boltzmann transport theory. The $\beta'$ phase is either the most stable
phase ($\beta'$-SnTe), or close to the most stable phases (such as orthorhombic phase of SnSe
and SnS) which are experimentally observed. Phonon dispersion relation calculation and
elasticity criteria are used to confirm the structural stability. A low lattice thermal
conductivity is obtained for $\beta'$-SnX, mainly because of hybridization acoustic phonon
modes with low-frequency inter-layer shearing vibration modes. A decent value of power factor
($\sim$ 0.01 Wm$^{-1}$K$^{-2}$) is also observed from our calculations, which is ascribed to
band convergence at CBM and quartic electronic band dispersion at VBM. A competitive dimensionless
figure of merit can be obtainable in $\beta'$-SnX within practical doping of a few 10$^{12}$
cm$^{-2}$, in particular, $ZT$ over 2.5 can be reached in $\beta'$-SnTe at 900 K. Thermoelectric
performance of $\beta'$-SnX can be further optimized with transport along zigzag crystalline
direction. Our theoretical study deems to facilitate discovering new phase for optimizing
thermoelectric performance by experiment.

\section{\label{sec:level1}ACKNOWLEDGEMENT}
This work is supported by the National Key R$\&$D Program of China (2017YFA0206301) and the
Major Program of Aerospace Advanced Manufacturing Technology Research Foundation NSFC and
CASC, China (No. U1537204). A.R. O. and B.J. D. thank the Russian Science Foundation (Grant
16-13-10459). R.S. acknowledges MEXT-Japan Grants Nos. JP25107005, JP25286005, JP15K21722
and JP18H01810. N.T.H. acknowledges JSPS KAKENHI Grants No. JP18J10151. Z.H. W. thanks the
National Science Foundation of China (Grant No.11604159)
and the Russian Scientific Foundation (Grant No.18-73-10135). Calculations were performed
on XSEDE facilities and on the cluster of the Center for Functional Nanomaterials, Brookhaven
National Laboratory, which is supported by the DOE-BES under contract no. DE-AC02-98CH10086.
This work has been carried out using the Rurik supercomputer, the Arkuda supercomputer of
Skolkovo Foundation, and computing resources of the federal collective usage center Complex
for Simulation and Data Processing for Mega-science Facilities at NRC Kurchatov Institute.

%

\end{document}